\documentclass[showpacs,
twocolumn,
prb,aps,eqsecnum,groupedaddress]{revtex4}
\usepackage{amsmath,amssymb,latexsym,amsfonts}
\usepackage[dvips]{graphicx}
\newcommand{\be}{\begin{equation}}
\newcommand{\ee}{\end{equation}}
\newcommand{\bea}{\begin{eqnarray}}
\newcommand{\eea}{\end{eqnarray}}

\begin{document}
%\draft
\title{Simulation of Electron Transport through a Quantum Dot with Soft Walls}
\author{Bernhard~Weingartner}
\author{Stefan~Rotter}
\author{Joachim~Burgd\"orfer}
\affiliation{Institute for Theoretical Physics, 
Vienna University of Technology,
Wiedner Hauptstr.~8-10/136, A-1040 Vienna, Austria, EU} 
\date{\today}
%%%%%%%%%%%%%%%%%%%%%%%%%%%%%%%%%
\begin{abstract}  
We numerically investigate classical and quantum transport through a 
soft-wall cavity with mixed dynamics. Remarkable differences to 
hard-wall quantum dots are found which are, in part, related to
the influence of the hierarchical structure of classical phase space on 
features of
quantum scattering through the device. We find narrow isolated transmission
resonances which display asymmetric Fano line shapes. The dependence of the
resonance parameters on the lead mode numbers and on the properties of
scattering eigenstates are analyzed. Their interpretation is aided by a
remarkably close classical-quantum correspondence. We also searched for
fractal conductance fluctuations. For the range of wave numbers $k_F$
accessible by our simulation we can rule out their existence. 
\end{abstract}
%\narrowtext
\maketitle
\section{Introduction}\label{sec:intro}
Theoretical investigations of ballistic transport through microstructures 
have shown that the 
spectral and transport properties of phase-coherent quantum systems, commonly 
called "billiards", depend strongly on the nature of the underlying classical 
dynamics.\cite{gutz91} 
To date, most investigations have focused on the two limiting cases 
of systems with either purely chaotic or regular dynamics. However, neither of 
these cases is generic.\cite{markus74} For the semiconductor 
quantum dots that are realized in the experiment\cite{reed95} 
a classical phase space structure with mixed regions of
chaotic regular motion is expected.
This is due to the fact that the boundaries of such 
devices are typically not hard walls (as in most theoretical investigations)
but feature soft wall
profiles for which such a "mixed" phase space is
characteristic.\cite{licht92} The investigation of quantum transport through
soft-walled microstructures is the primary goal of the present
communication.\\The mixed classical phase space results in specific transport
properties. Consider, e.g., the classical escape rate from an open
billiard. For a dot with ``hard chaos'', i.e.~a metrically transitive system,
the dwell time distribution, or equivalently the length distribution $P
(l)$ decays exponentially, $P (l)\sim e^{-l/\bar{l}}$ (with $\bar{l}$
being the mean path length). In a mixed
system, however, trajectories can be trapped in the vicinity of regular
islands, leading to an increased length distribution which typically
features an algebraic decay law, $P(l) \sim (l/\bar{l})^{-\alpha}$. In those
hard-wall billiards with shapes that allow for a mixed phase 
space, it was shown that trapped trajectories lead to quasi-bound states
in the corresponding quantum
transport problem and appear as isolated resonances
in the conductance.\cite{hucke00,hucke00e,baeck02} 
In the chaotic-to-regular crossover regime also so-called
``Andreev-billiards''\cite{henning,libisch} and the effect of
shot noise suppression\cite{sim02,aigner,sukho} have recently been
discussed. Furthermore, 
mixed classical dynamics was proposed as a mechanism giving 
rise to fractal conductance fluctuations (FCF).\cite{ketzprb96} 
Several experiments have meanwhile been performed to
test this prediction. First experimental
data\cite{sachr98,micolich01,micolich02,micolich04,crook03} 
appear to support this
notion. However, in the corresponding numerical studies, no fractal structure
in the conductance fluctuations could be
found,\cite{hucke00,hucke00e,baeck02,taka00} which, in part, has led to a
number of theoretical works that propose alternative and sometimes even
contradictory explanations for
FCF.\cite{guarneri01,benenti01,budiyono03,louis00} 
In the present paper we inquire
into the appearance of this features for transport through soft-walled quantum
dots.\\\linebreak
We calculate transport 
coefficients and scattering wave functions by
solving the time-independent one-particle Schr\"odinger equation for a
two-dimensional scattering device. Of particular interest is the semiclassical
limit
of transport, where the Fermi wavelength 
$\lambda_F=2\pi/\sqrt{2E_F}$ (in a.u.) is much smaller
than the linear dimension $D$ of the quantum billiard, $\lambda_F\ll D$.
This is because the ratio $(\lambda_F/D)$ determines the resolution with which
quantum mechanics can resolve the underlying (mixed) classical phase
space. Note, however, that the semiclassical limit for transport in the leads
of
width $d,\ \lambda_F \ll d$, cannot be reached. The asymptotic incoming
and outgoing scattering states thus 
remain in the quantum regime. In the limit of
large Fermi energies $E_F$ quantum transport simulations are quite
demanding. In order to reach the high energy regime we employ the {\it Modular
  Recursive Green's Function Method} (MGRM),\cite{rotter00,rotter03} which is
a variant of the standard recursive Green's 
function approach\cite{akisprl97} suited for small
wavelength.\\For a detailed analysis of 
the influence of the mixed classical phase space on the quantum
scattering problem we compare the classical Poincar\'e surface of section with
the Husimi distribution derived from the scattering wavefunctions.
We are thereby able to classify the isolated
conductance resonances and find that
a recently suggested classification\cite{baeck02} in terms of 
scattering states corresponding to classically regular or trapped
trajectories, has to be extended to include  
resonances which are associated with unstable periodic orbits.
The quantum counterpart of these orbits (commonly called
``scars''\cite{hellerprl53}) emerge in the wavefunction densities
which we calculate numerically.
These findings are supported by very recent experimental investigations of a
soft-wall microwave
billiard for which scarred wavefunction have been, indeed, observed.\cite{kim} 
We demonstrate that the resonances in conductance follow the characteristic 
asymmetric Fano lineshape\cite{fano61} and perform a statistical
analysis of the distribution of resonance widths, amplitudes and Fano
asymmetry parameters. Finally, we inquire into the occurrence of fractal
conductance
fluctuations.\\This paper is organized as follows. 
In Sec.~\ref{sec:classical} we present the scattering 
device investigated in this work. Section~\ref{sec:isolated} is dedicated to a 
discussion of the isolated conductance resonances and the 
corresponding wavefunctions. The Fano profile of the resonances is analyzed in 
Sec.~\ref{sec:fano}. In Sec.~\ref{sec:fractal} we discuss fractal conductance
fluctuations and Sec.~\ref{sec:summary} finally gives a summary of the
results. 
\section{Classical dynamics}\label{sec:classical}
\begin{figure}[!b]
\includegraphics[width=85mm]{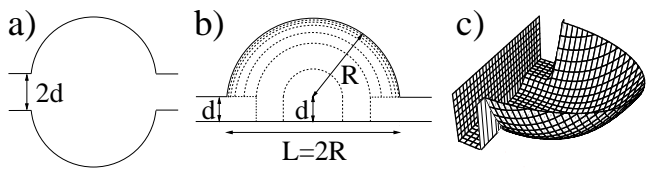}
%{pictures/geom.alg.mod.pot35.pstex}
 \caption{(a) Bunimovich stadium with leads attached at the 
   straight segments. (b) Reduced scattering geometry consisting of 
a semi-circle of radius $R=1$ which is attached to a rectangle of length $L=2R$
and width $d=0.3$. The solid lines represent hard-wall boundary conditions and
the
dashed curves are contour lines of the soft-wall potential [for details see
Eq.~(\ref{eq:2.3})]. (c) Illustration of the potential 
surface corresponding to contour plot (b).}
\label{fig:1}
\end{figure}

\begin{figure}[!t]
\includegraphics[width=85mm]{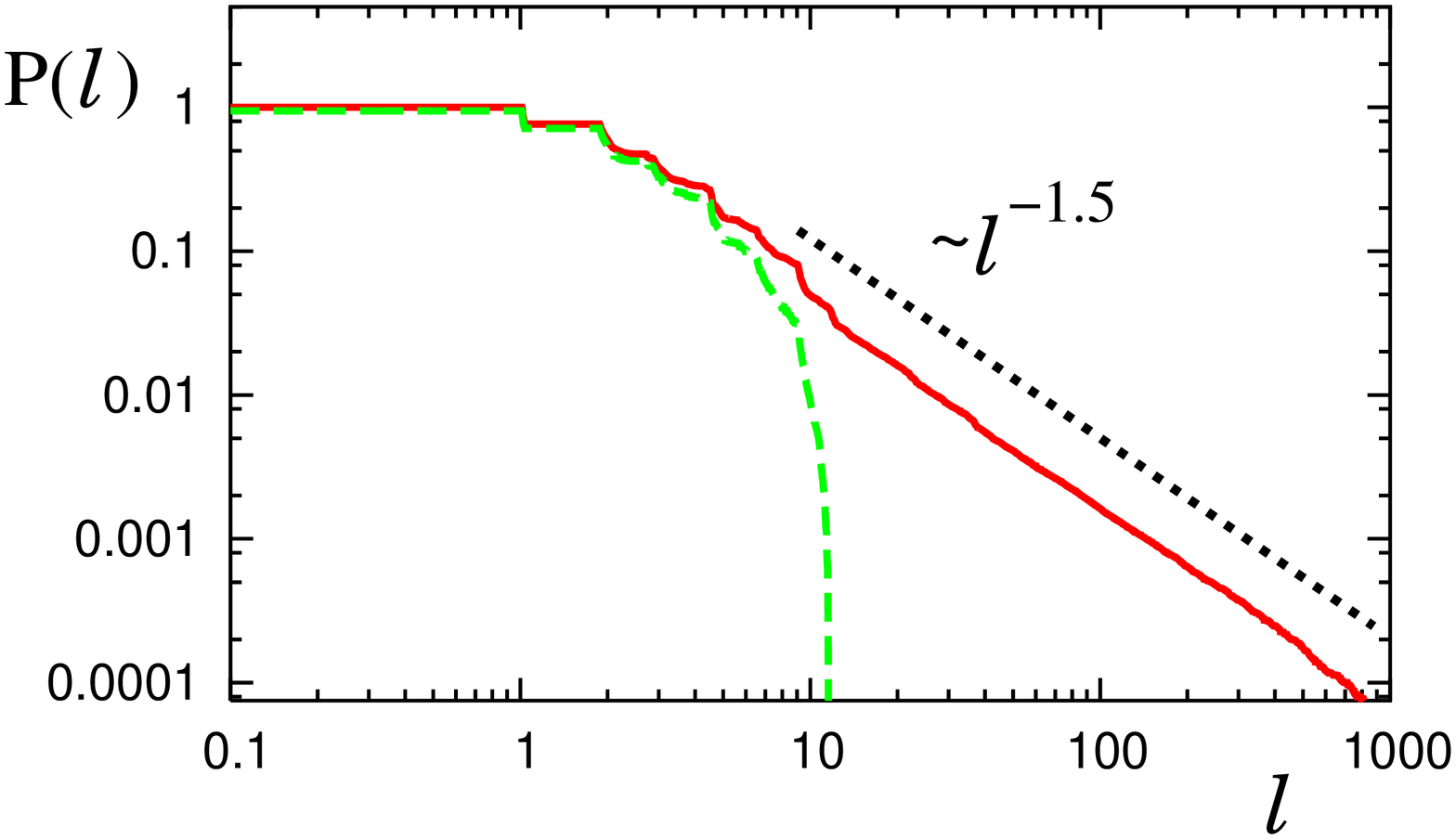}
%{pictures/geom.alg.mod.pot35.pstex}
 \caption{(Color online) 
   Log-log plot of the classical dwell probability $P(l)$ for the
   cavity with soft walls (red solid
   line) and hard walls (green dashed line).}
\label{fig:2}
\end{figure}

\begin{figure}[!b]
\includegraphics[width=85mm]{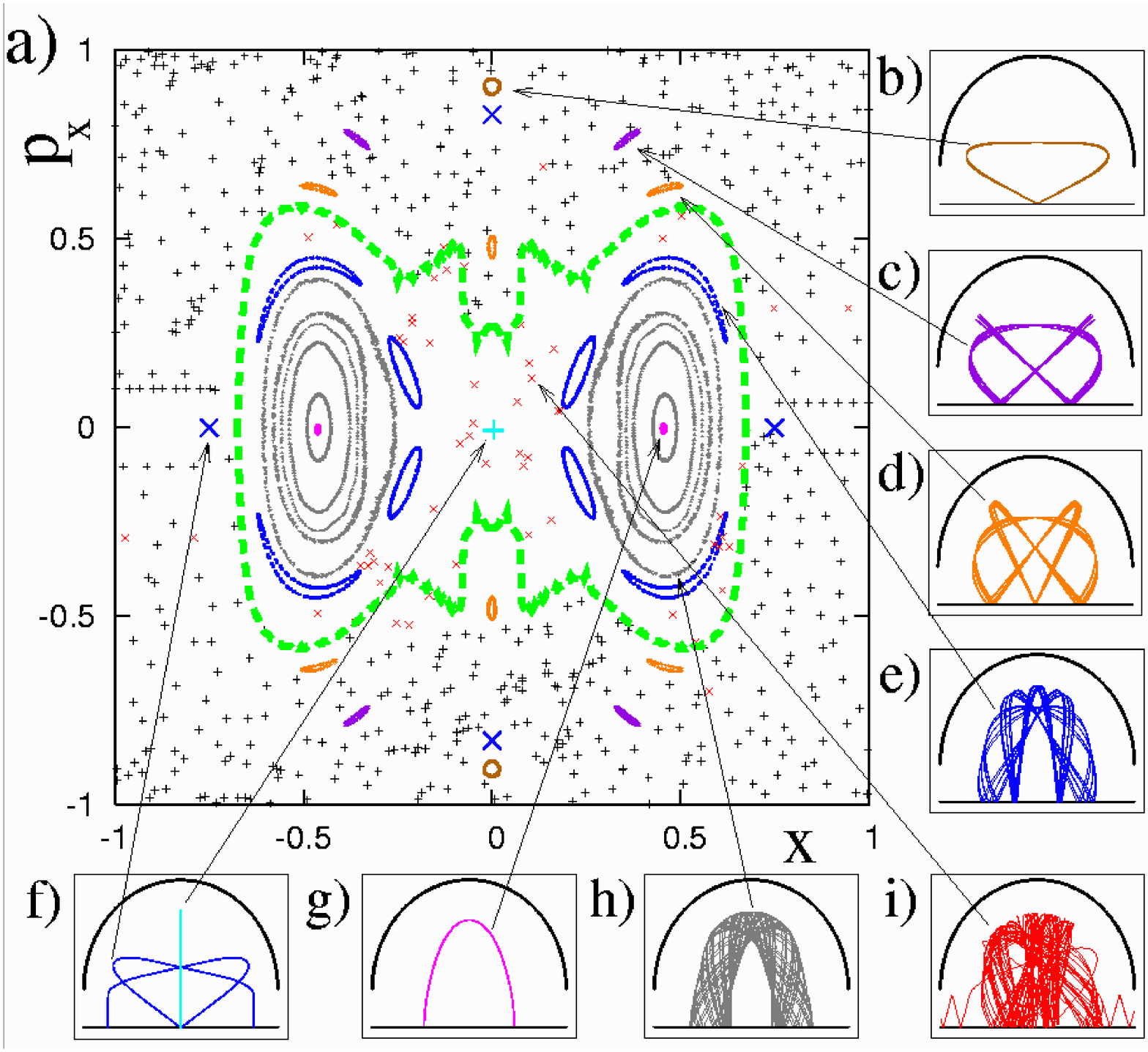}
 \caption{(Color online) Poincar\'e surface of section for the
half-stadium with soft walls (see Fig.~\ref{fig:1}b,c). A large
number of chaotic trajectories (black crosses), six
regular closed orbits (b-e),(g),(h), two unstable periodic orbits
(f) and one trapped chaotic trajectory
(i) are shown. For orientation the arrows denote one point of the 
phase space representation of each trajectory. The green dashed curve shows
an approximation for the
outermost partial transport barrier. For details see text.}
 \label{fig:3}
 \end{figure}

We consider in the following a stadium-shaped quantum dot with
a semicircle of radius $R$ and leads attached to the straight section (see
Fig.~\ref{fig:1}a). 
By focusing on scattering states with odd parity under
reflections $(y \rightarrow  -y)$, the geometry can be reduced to one
semicircle with attached leads with half of the original width. The 
resulting cavity boundary, including the added
soft-wall potential, is depicted in
Fig.~\ref{fig:1}b,c. 
We inject electrons from the left into this system and study the transmission
and reflection probabilities classically as well as quantum mechanically.
Classical simulations are performed by calculating many different trajectories
of electrons with Fermi energy $E_F$ 
which enter the dot at $x=-R$. The initial positions
across the lead width are uniformly distributed with an 
angular distribution  $P(\theta)\sim \cos(\theta)$.
The ballistic quantum scattering problem is solved with the 
{\it Modular Recursive Green's Function Method} 
(MRGM).\cite{rotter00,rotter03} 
We calculate scattering wave functions
and the $S$ matrix of the system at different Fermi energies $E_F=k_F^2/2$ (in
a.u.), where $N={\rm int}(d\,k_F/\pi)$ modes are transmitting in the leads. 
The total transmission $T$ is then given by
\begin{equation}
T=\sum^N_{m=1}T_{m}=\sum^N_{m,n=1}T_{mn}=\sum^N_{m,n=1}|t_{mn}|^2
\label{eq:2.1}
\end{equation}
with $t_{mn}$ being the transmission amplitudes from incoming mode $m$ to
outgoing mode $n$. According to the Landauer formula, the conductance is obtained as
\begin{equation}
G = 2\times T \, .
\label{eq:2.2}
\end{equation}
Atomic units ($\hbar=|e|=m_{\rm eff}=1$) 
will be used unless otherwise stated explicitly.\\If 
we choose a zero potential inside the structure and hard-wall boundary
conditions, the scattering device corresponds to the open Bunimovich stadium
billiard which is prototypical for purely
chaotic dynamics.\cite{bunimo74,benettin78}
This behavior changes drastically if a soft wall profile is introduced 
(see dashed contour lines in Fig.~\ref{fig:1}b and Fig.~\ref{fig:1}c), 
given by:
\begin{equation}
\phantom{!}\label{eq:2.3}
\end{equation}\vspace*{-1.4cm}
\begin{eqnarray}
\lefteqn{V (\vec{r}\;\!)=}\nonumber\\
&&\phantom{+}\ \Theta (d-y)\left\{
\begin{array}{c@{\quad}l}
-A, & |x| < 0.35\\ {\rm min}[0,B(|x|\!-0.35)^2 -A], & |x| > 0.35 
\end{array}
\right.
%\ {\rm min} \left[ 0, \left( -A + B (x - 0.35)\right)^2 \right] 
\nonumber\\&& \nonumber +\ \Theta (y - d) 
\left\{
\begin{array}{c@{\quad}l}
-A, & \phantom{|}S\phantom{|} 
< 0.35\\ \phantom{\,{\rm min}[0,}B(S\phantom{|}\!
-\phantom{|} 0.35)^2 -A\phantom{]}\!, & \phantom{|}S\phantom{|} > 0.35 
\end{array}
\right.\phantom{+++++}
\end{eqnarray}
with $S = |\vec{r} - \vec{d}_0 |$ and $\vec{d}_0 = (0, d)$.
$V (r)$ rises quadratically in the exterior region of the semicircle. 
The potential in the rectangular region below the half-circular module depends
only on the $x$-coordinate. It follows the radial profile of $V(r)$, however,
only for potential values below the equipotential line of the lead such that
injection and emission is barrier-free. Note that in contrast to the case
where all boundaries are hard walls, the classical scattering dynamics with an
arbitrary soft-wall profile is not invariant under scaling of the electron
energy $E_F$. In order to approach the semiclassical limit of pathlength
spectroscopy, we choose a scaled potential with $ A = A_0\times E_F$ and $B =
B_0\times E_F$. This results in classically scaling invariant dynamics and all
quantum results we obtain can be compared with one and the same classical
phase space structure.\\As a first test of the effect the soft-wall potential
has on transport, we plot
in Fig.~\ref{fig:2} the probability distribution $P(l)$ for classical
trajectories to leave the cavity after a length $l$. 
We find for the
chaotic case with hard walls an exponential decay\cite{lin93,rotter00} of
$P(l)$ and for the mixed case due to the soft walls a power-law behavior
which is well approximated by $P(l) \sim l^{-1.5}$ over several orders
of magnitude. The difference to the chaotic case can be understood as a
signature of the trapping of trajectories in the vicinity of the hierarchical
set of regular islands which is typical for mixed 
dynamics.\cite{meiss86,ketzprb96} For a more detailed phase space analysis we
plot the Poincar\'e surface of section (PSS) of the classical dynamics in our
soft-wall device. At each bounce of a trajectory against the horizontal lower
boundary, the position $x$ along the boundary and the projection of the
momentum vector in the horizontal direction $p_x$ are
recorded. Fig.~\ref{fig:3} shows
the PSS for an ensemble of initial conditions with several islands of regular
motion and unstable periodic orbits in an otherwise chaotic sea. Of particular
interest are trajectories trapped in the vicinity of the islands (see e.g.,
Fig.~\ref{fig:3}i). 
The trapping region of phase space is shielded from the
surrounding
chaotic phase space by partial transport barriers which are formed by cantori
as well as by stable and unstable manifolds.\cite{licht92} A partial barrier
can only be crossed through small gaps (``turnstiles'') where phase space
volume is exchanged.\cite{mackay} Trajectories that enter the cavity through
the entrance lead first reach the chaotic part of phase space and may directly
exit through the exit lead. Some trajectories do, however, cross the outermost
partial barrier through a turnstile and stay trapped inside for a
comparatively long time since the only path to the exit is via another
or the same turnstile. Alternatively, these trajectories may get
trapped
even deeper inside the next layer of the hierarchical set of transport
barriers. By contrast, the islands of regular motion which lie at the core of
this hierarchy are invariant curves forming complete barriers and can thus not
be accessed by classical trajectories emanating from the leads and
contributing to transport.\\In order to determine the shape of the partial
barriers and the location and
size of the turnstiles the analytical methods described in
Ref.~\onlinecite{mackay}
could, in principle, be used. It is, however, quite demanding to construct the
partial barriers explicitly, in particular in the present case of soft-wall
cavities where analytical solutions for trajectories are not available. We
have therefore determined the outermost barrier approximately by a numerical
dwell length analysis. We scan the dwell length $l$ inside the cavity on a
very dense array of points in the PSS. All initial conditions which correspond
to a dwell length longer (shorter) than a typical threshold value, $l_0 =
15$, are assumed to be inside (outside) the outermost partial barrier. As a
result we obtain the approximate partial barrier line depicted in the PSS of
Fig.~\ref{fig:3} as a
green dashed line. The location of the barrier was found to be
only very weakly dependent on the choice of $l_0$.
\section{isolated resonances}\label{sec:isolated}
\begin{figure}[!t]
\includegraphics[width=85mm]{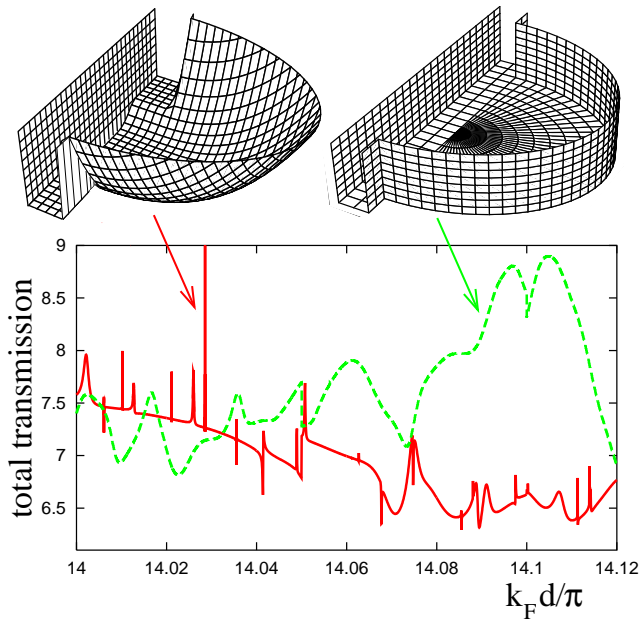}
\caption{(Color online) Comparison between total transmission $T$ through the
  soft-wall (left inset) and hard-wall geometry (right inset)
as a function of the Fermi wavenumber $k_F$ (14 lead modes are open).
Narrow isolated resonances on a smoothly varying background are clearly
visible for the soft wall billiard (red solid line), 
but are absent for the case of a hard wall boundary 
(green dashed line).}
\label{fig:4}
\end{figure}

\begin{figure}[!b]
\centering
\includegraphics[width=85mm]{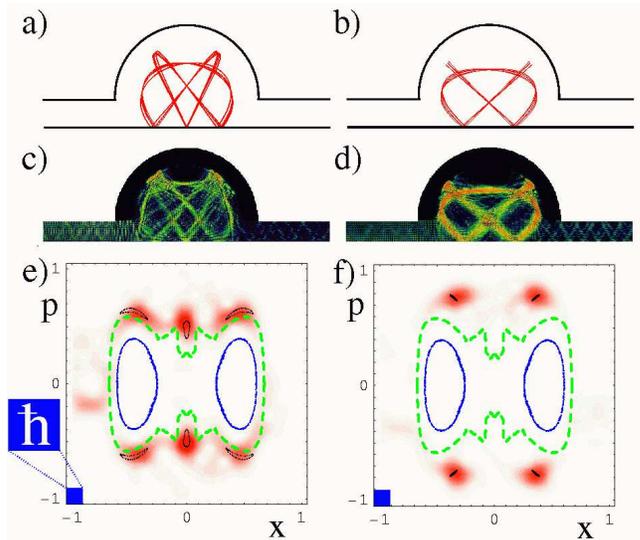}
\caption{(Color online) (a),(b) Classical periodic orbits in the soft-wall
  billiards
which are closely mirrored in the density $|\psi(x,y)|^2$ of the quantum
scattering
wavefunctions at the resonant energies: (c) wavenumber $k_F=14.0592\pi/d$ for
incoming lead mode number $m=14$ and (d) $k_F=14.0021\pi/d,$ with
$m=12$\,. Figure (e) and (f) give the corresponding phase space
portraits: Husimi distribution (density plot, red) and 
classical PSS (islands, black). For reference we also plot the outermost
partial barrier (green dashed line) and the most prominent regular island 
(blue solid lines). 
The blue boxes in the lower left corner of (e),(f) indicate
the size of $\hbar$ which characterizes the quantum mechanical resolution of
the classical phase space.}
\label{fig:5}
\end{figure}

\begin{figure}[!ht]
\includegraphics[width=85mm]{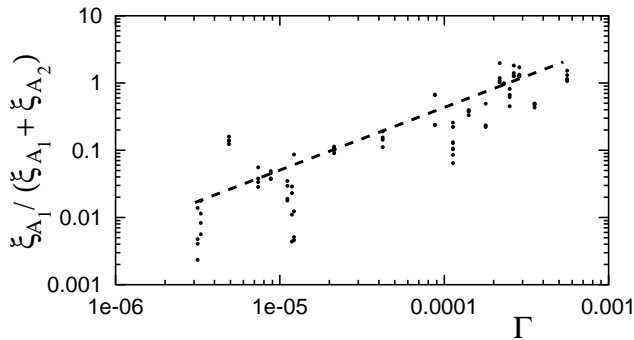}
\caption{The ratio $\xi_{A_1}/(\xi_{A_1}+\xi_{A_2})$ of the HDs integrated
  over the 
    chaotic area $A_1$ outside the outermost partial barrier and
    over the remaining phase space area $A_2$ [see 
    Eq.~(\ref{eq:3.2})] as a function of the resonance width
    $\Gamma$.}
\label{fig:7}
\end{figure}

Quantum dynamics profoundly modifies the phase flow in the presence of cantori
in two ways: On the one hand, partial barriers become impenetrable when the
phase space volume of the turnstile is smaller than that of the Planck cell
$(\sim \hbar)$ i.e.~the size of the minimum-uncertainty
wavepacket. Consequently, phase flow is suppressed on a short time scale
associated with classically allowed transitions. On the other hand, quantum
mechanics opens up the possibility of barrier penetration of both complete and
partial barriers by tunneling. This purely quantum transport channel is,
however, in general slow and associated with the time scale for
tunneling. Thus, the outermost partial barrier with turnstiles smaller than
$\hbar$ divides the phase space into two distinct regions where chaotic and
hierarchical eigenfunctions are concentrated on either side.\cite{ketzprl00}
Since the hierarchical region couples only very weakly to the leads,
quasi-bound states residing in this part of phase space give rise to sharp
resonances in transmission.\cite{hucke00,hucke00e,baeck02} 
Classically, an island of
regular motion in phase space consists of a set of concentric invariant
curves, each of them forming a complete barrier. Accordingly, long-lived
quasi-bound states reside also in islands of regular motion and account for
additional narrow resonances in transmission.\\In Fig.~\ref{fig:4} we compare
the conductance (i.e.~transmission) through our
device as a function of the Fermi wavenumber $k_F$ for soft-wall and hard-wall
potentials (see insets). 
Note the remarkable difference between the two
graphs: Sharp resonances are present in the case with a mixed phase space (see
Fig.~\ref{fig:3}) 
and completely absent for the chaotic case. The quantum counterpart to
the classical PSS is the quantum phase space
distribution. In the following we analyze the Husimi distribution $H(x,p)$ at
the resonance energy and investigate its localization.\cite{baeck02} We
define the Husimi distribution (HD) in direct analogy to the classical PSS by
projection of the scattering state onto a coherent state $\phi^{\rm coherent}$
on the lower horizontal boundary.\cite{baeck04} The HD reads
\begin{eqnarray}
H(x,p)&=&|\langle\partial_{\bf n}\psi(x,y)|\phi^{\rm
  coherent}(x,p)\rangle|^2=\\
&=&\left|\int_{-1}^{1}dx'\partial_{\bf n}\psi^*(x')e^{ik_Fp(x'-x)-(1/2)k_F(x'-
x)^2}\right|^2\,,\nonumber
\end{eqnarray}
where $\partial_{\bf n}\psi(x)={\bf n}(x)\cdot\nabla\psi(x,0)$ is the normal
derivative 
of the scattering state on the lower boundary and ${\bf n}(x)$ is the vector
normal to
 the lower cavity wall.\\In Fig.~\ref{fig:5} we show both wavefunctions and
 Husimi distributions for
 two resonant scattering states for which the quantum-classical correspondence
 between classical trajectories (a)-(b) and the scattering wavefunction
 (c)-(d) is particularly striking. 
 Correspondingly, the HDs of these
 scattering states reside on top of the regular islands within which these
 periodic classical orbits propagate.\\This remarkable degree of
 quantum-classical correspondence allows a convenient characterization of
 resonances in terms of the underlying phase space structure. The observed
 resonances fall into three classes: Resonances that are associated with the
 regular $(R)$ or with the hierarchical 
 $(H)$ regions in phase space, 
 and finally those that are associated with scars $(S)$, i.e.~unstable
 periodic orbits in the chaotic sea. $R$ and $H$ resonances have also been
 found in hard-walled billiards with boundaries that allow for a mixed phase
 space\cite{baeck02} while scars are well-known features in bound-state
 wavefunctions of closed metrically transitive  (``hard chaos'')
 systems.\cite{hellerprl53} We note that $S$ resonances have been very
 recently identified in soft-walled 
 microwave billiards.\cite{kim} Typical examples are
 shown in Fig.~\ref{fig:6}: 
\begin{figure*}
\includegraphics[width=170mm]{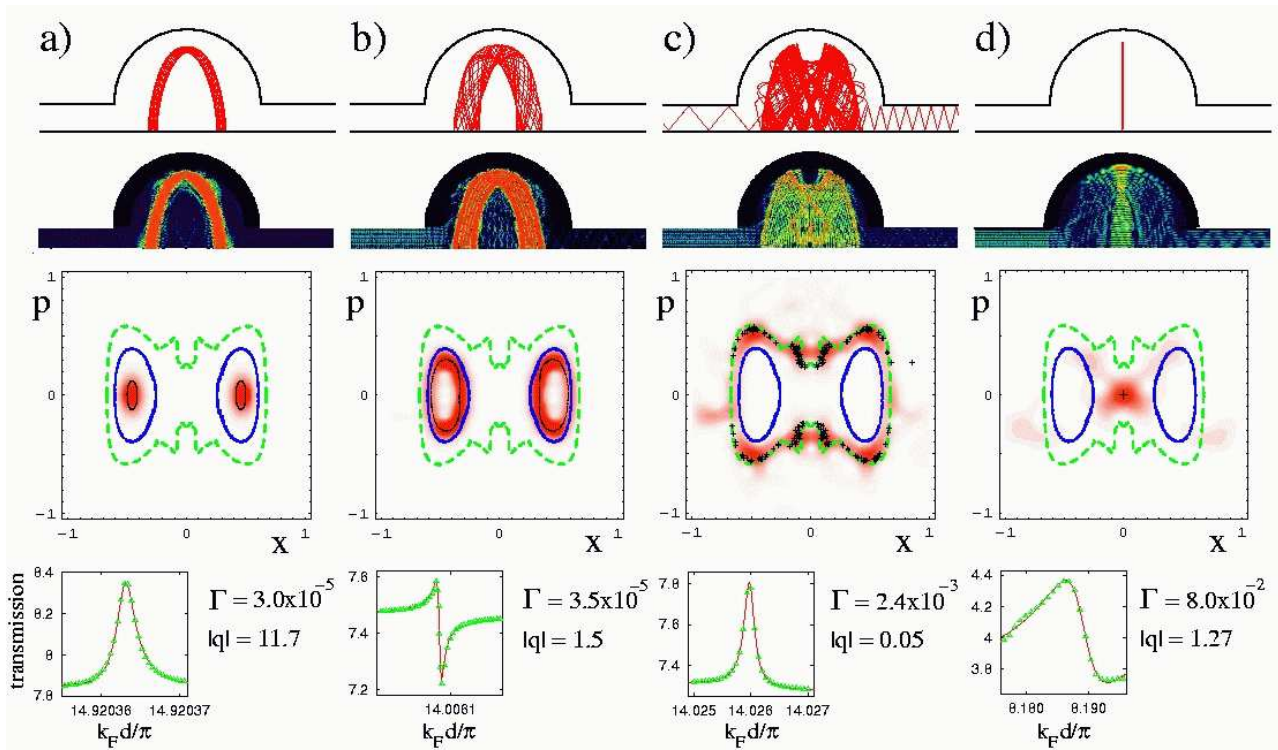}
\caption{(Color online) Top row: Classical stable periodic orbits (a) and (b);
a transiently trapped trajectory (c) and an unstable periodic
orbit (d). Second row: Resonant wave function densities $|\psi(x,y)|^2$
closely mimicking these
trajectories with wavenumbers (a) $k_F=14.9203\pi/d$, (b) $k_F=14.0061\pi/d$,
(c) $k_F=14.0260\pi/d$, and (d) $k_F=8.186\pi/d$. Third row: The Husimi
distributions (density plot, red) and the PSS of the corresponding
trajectories (island and crosses, black). 
Bottom row: Transmission resonances, numerical results (green
dots) and fits to the Fano resonance formula Eq.~(\ref{eq:4.1}) (red
line). Fitted values for the width $\Gamma$ and Fano asymmetry parameter $q$
are indicated.}
\label{fig:6}
\end{figure*}
 Resonances (a) and (b) clearly fall into the class
 $R$, for which the HD resides on regular islands. State (a) corresponds to
 the center of the two prominent stable islands whereas state (b) resides near
 the outer border of the island. The corresponding resonance widths $\Gamma$
 mirror this difference: Although both being narrow and of the same order of
 magnitude, $\Gamma \approx 10^{-5}$, the width decreases from the 
 state near the outer border
 (b) to the state in the center of the island (a), indicating that the
 probability to tunnel out of the island is higher near the border than from
 the center of the island (for quantitative details see
 Fig.~\ref{fig:6}). Note that these results are in close correspondence with
 the findings presented in Ref.~\onlinecite{ketzprl05}.\\The HD of a
 ``hierarchical'' state (c) features
 pronounced intensity in the region near the cantorus corresponding to a
 classically chaotic trajectory that gets transiently trapped inside the
 partial barrier. Compared to the regular states the coupling of the
 hierarchical states to the leads is stronger and therefore
results in a resonance width
 $\Gamma \approx 10^{-3}$ which is typically two orders of magnitude larger
 than for the regular states. State (d) corresponds to a ``scarred''
 wavefunction whose classical analogue is an unstable periodic
 ``bouncing-ball'' orbit. Its resonance width $\Gamma  \approx 10^{-1}$ is
 even larger than of most of the hierarchical states we recorded. We note,
 however, that the very limited number of such scarred states which we could
 identify
 prevents us from drawing definite conclusions about the generic value  of
 their width. Nevertheless, it is reasonable to guess that
 the relation found for the present system $\Gamma_R \ll \Gamma_H \ll
 \Gamma_S$ should hold in other systems as well.\\ The close classical-quantum
 correspondence in the phase structure suggests that the width of the
 transmission
 resonances can be estimated from the location of the HD relative to the
 cantorus. Specifically, we decompose the HD into one part that lies outside
 the cantorus occupying the area $A_1$ in the PSS and the complementary area
 $A_2$ lying inside the cantorus.\\To quantify the overlap of the HD with these
 areas we integrate the HD according to
\begin{equation}
\label{eq:3.2}
\xi_A=\int\!\int_A dx\,dp\;H(x,p)\,.
\end{equation}
In Fig.~\ref{fig:7} the ratio $\xi_{A_1}/(\xi_{A_1}+\xi_{A_2})$ is plotted for
each resonance as a function of the resonance width $\Gamma$ for a large
number of resonances which
we analyzed. 
We find the proportionality 
\begin{equation}
\label{eq:3.3}
\Gamma \propto \left[ \xi_{A_1} / (\xi_{A_1}+
\xi_{A_2}) \right]^{1.27\pm 0.12}\, .
\end{equation}
Within the statistical uncertainty the deviation from linearity is most likely
not significant. An approximately linear dependence on 
$\xi_{A_1} / (\xi_{A_1} +
\xi_{A_2})$ could be expected for $H$ resonances and for those $R$ resonances
corresponding to islands inside the cantorus. Clearly $S$ and the
remaining $R$
resonances fall outside the validity of this estimate. 
\section{Fano profile} \label{sec:fano}
\begin{figure}[!t]
\includegraphics[width=85mm]{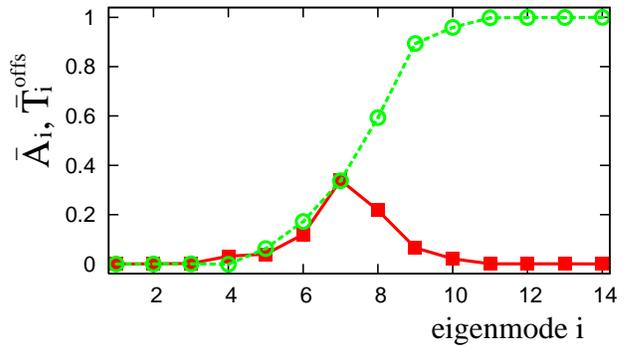}
\caption{(Color online) Offset of transmission eigenvalues 
$\overline{T}^{\rm \,offs}$ (green dashed line, open circles) and 
amplitudes $\overline{A}_i$ (red solid line, squares) 
of the Fano resonances in the eigenmode $i$.
All Fano resonances in the ensemble considered were
taken from the energy interval where 14 lead modes are open, resulting
in $i_{\rm max}=14$. The resonance parameters associated
with the eigenmode $i$ (for ordering see text) are ensemble averaged  
within one specific eigenmode $i$. Note that the 
amplitudes $\overline{A}_i$ feature pronounced values only for
intermediate modes $i$ where the resonance offset
$\overline{T}^{\rm \,offs}$ does not take on the ``classical'' values close to
$0$ or $1$.}
\label{fig:9}
\end{figure}

\begin{figure}[!ht]
\includegraphics[width=85mm]{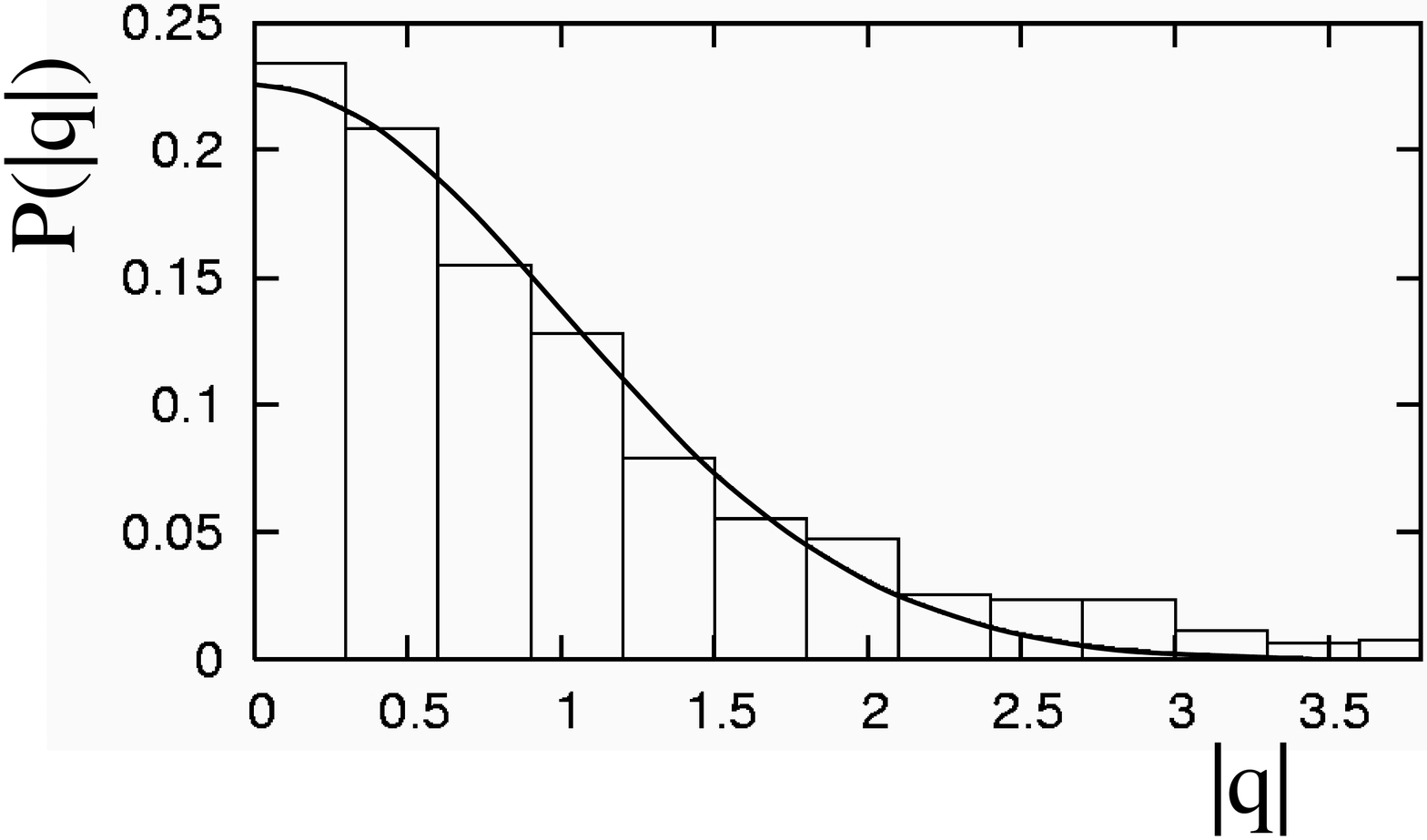}
\caption{Distribution $P(|q|)$ of the absolute value of the 
$q$-parameter in Fano resonances found in the
transmission eigenvalues $\lambda_i$ (the same ensemble of resonances was
considered as in Fig.~\ref{fig:9}). Fitting the data (histograms) 
with a Gaussian distribution yields good agreement.}
\label{fig:10}
\end{figure}

The isolated narrow resonances in transmission $T(k_F)$ have typically an
asymmetric Fano line shape\cite{fano61} illustrated in the bottom row of
Fig.~\ref{fig:6}. Fano resonances have been observed in many different fields
of physics, including ballistic transport through quantum
dots.\cite{rotter03,noeckel94,goeres00,rotter04,kobayashi02,condmat04} 
They occur when
(at least) one resonant and one non-resonant pathway connecting the entrance
with the exit channel interfere. The specific interest in analyzing Fano
profiles is driven by their high sensitivity to the details of the scattering
process, in particular the degree of coherence in transport and
the presence of decoherent interactions with other degrees of freedom. In
contrast to Breit-Wigner resonances, the asymmetric Fano resonances are not
only determined by their resonance position $k_R$, width $\Gamma$ and
amplitude $T_0$, but also by the Fano asymmetry parameter $q$ according to
\begin{equation}
\label{eq:4.1}
T(k_F)  \approx T^{\rm offs}+T_0\times\frac{(k_F - k_R +
q\,\Gamma/2)^2}{\left(k_F-k_R\right)^2+\left( \Gamma/2 \right)^2}\,,
\end{equation}
where $T^{\rm offs}$ is the offset value of the resonance minimum.
Note that the smoothly varying background on top of which the resonance
is situated is thus given by: $T^{\rm bgr}=T^{\rm offs}+T_0$\,.
The relative amplitude $A$ of the resonance (i.e.~the difference between
maximum 
and minimum value of the second term in Eq.~(\ref{eq:4.1})) is given
by\cite{fang} $A=T_0\times(1+q^2)$. 
Due to the time-reversal symmetry in our system (i.e.~no magnetic
field or decoherence present) the asymmetry
parameter $q$ can be treated as real and is a measure for the ratio 
between resonant and non-resonant transmission amplitude.\cite{fano61} For 
$|q|\rightarrow0$ non-resonant transmission dominates resulting in a symmetric 
dip at the resonant position. For $q\approx 1$ the peak is highly
asymmetric. In the absence of non-resonant transmission, i.e.~$|q| \rightarrow
\infty$, the resonance shape approaches that of a Breit-Wigner profile. For
coherent transport in the low-energy regime, where only one flux-carrying mode
is open, the transmission will vary between its maximum value $T(k_F)$ = 1
(full transmission) and $T(k_F)=0$ near each resonance in the single-mode
limit. This implies $T^{\rm offs}=0$ and 
$A=1$ in Eq.~(\ref{eq:4.1}). As soon as $k_F$
passes the threshold $k_F = 2 \pi/d$ for opening up additional transmitting
modes, Fano resonances have, in general, a minimum different from zero,
$T^{\rm offs}\neq 0$.% and feature relative amplitudes $A<1$.
\begin{figure}[!b]
\includegraphics[width=85mm]{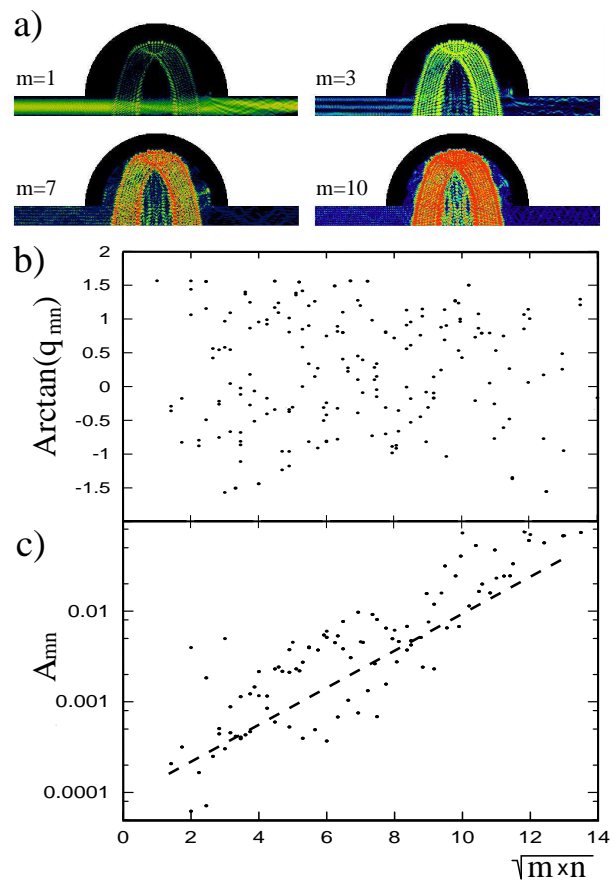}
\caption{(Color online) (a) Wavefunction densities for selected incoming mode
numbers $m$ of a typical Fano resonance [$k_F=14.1111\pi/d$].
For increasing $m$ the directly transmitted (non-resonant) part of the
wavefunction, being
dominant for $m=1$, becomes less pronounced, while the relative
intensity of the $\cap$-shaped resonant part drastically increases. (b) 
The Fano
parameter $q$ shows no correlation to the geometric mean value
of the in- and outgoing mode number, $\sqrt{m\times n}$. (c) 
Amplitudes $A_{mn}$  are approximately proportional 
to $\sqrt{m\times n}$.}
\label{fig:8}
\end{figure}\\In order to
elucidate
the formation of Fano resonances we decompose the total
transmission $T(k_F)$ in terms of its contributions from different
modes.\cite{goldberger} A Fano resonance observed in $T(k_F)$ appears also as
a Fano resonance in the channel transmission probabilities
$T_{mn}$ at an identical position $k_R$ and width $\Gamma$. In general, the
values for $q$ and $T_0$ are however different in all the $m\times n$
channels. Conversely, the sum of any number of Fano resonances with identical
$\Gamma, k_R$ will again be a Fano resonance. Using the semiclassical
connection between mode number and injection angle, $\sin\theta =
m\pi/(dk_F)$, a naive expectation would be that
for the billiard geometry of Fig.~\ref{fig:1} and low
mode numbers $m,\,n$ a large fraction of transmission is mediated by direct
(non-resonant) transmission and only a small part by resonant transport which
corresponds to transient trapping inside the structure. Low-mode numbers in
the leads correspond classically to small injection and ejection angles which,
in our specific geometry, is equivalent to trajectories that
connect the entrance and exit lead without exploring much of the cavity. On
the other hand we expect resonant trapping
to dominate for large mode numbers $m,n$. 
This feature is illustrated in
Fig.~\ref{fig:8}a, where we plot the density of scattering wave functions with
the same
$k_F$ but different incoming mode number $m$. The resonant part of the wave
function  (in form of the $\cap$-shape) is strongly suppressed for the
limiting case $m = 1$ but increasingly pronounced for growing $m$. It is now
of interest\cite{ihra02} 
to explore the dependence of the two parameters, the partial
amplitude $A_{mn}$ and the Fano parameter $q_{mn}$ on the ``geometric''
variable $m\times n$. The lower limit
$\sqrt{m\times n}=1$ corresponds to (almost) horizontal injection and ejection
angles
with predominant non-resonant transport (see Fig.~\ref{fig:8}a, $m=1$) whereas
increasing 
mean values of the mode numbers stand for larger angles and therefore rising
dominance of the resonant pathway. 
While $A_{mn}$ is approximately proportional to $m\times n$ (see
Fig.~\ref{fig:8}c) indicating that with large injection and ejection angle the
relative
amplitude increases, we find the remarkable result that the values $q_{mn}$
are virtually uncorrelated to $m\times n$ 
(Fig.~\ref{fig:8}b), contrary to a naive picture. 
Intuitively, one would expect the asymmetry parameter $q$,
being a measure for the ratio 
between resonant and non-resonant contributions to transmission, to
systematically increase with $\sqrt{m\times n}$,
but no signs of 
proportionality between $\sqrt{m \times n}$ and $q$ can be detected in
Fig.~\ref{fig:8}b.\\The projections of the resonant states onto the lead
wavefunction are, however, not a basis invariant measure for transport. We
therefore explore the transmission eigenvalues, i.e.~the eigenvalues of the
transmission operator $\hat{T}=t t^\dagger$ with matrix elements
 \begin{equation}
 \label{eq:4.2}
 \langle n|T|m\rangle = \sum^{M}_{k=1} \langle n|t|k\rangle \langle 
k|t^\dagger | m\rangle
 \end{equation}
and $n, m, k \leq M$ corresponding to the number of open modes. The
transmission eigenvalues $\lambda_i$ ($i=1,\ldots,M$)\,, which we label in
ascending order ($\lambda_i\leq \lambda_{i+1},\,\forall i=1,\ldots,M-1$),
provide a channel-basis invariant representation. Implicit in this
analysis is the assumption that the matrix diagonalizing $T$ is only
weakly energy dependent across the width of the resonance. This is
justified for narrow and non-overlapping resonances, as we have verified
numerically in a few cases. We point out the similarity of this approach
to the multi-channel quantum defect theory employed in atomic and molecular
physics.\cite{seaton,fritz} We explore now the
appearance of Fano resonances in $\lambda_i$\,. For this purpose we determine 
$\lambda_i$ for a large number of Fano resonances and extract from each 
$\lambda_i$ the offset $T^{\rm offs}_i$, the amplitude $A_i$ and
the eigenchannel Fano parameter $q_i$. The corresponding averages over the
ensemble of Fano resonances are denoted by $\overline{T}^{\rm \,offs}_i$
and $\overline{A}_i$. As is well known from 
random matrix theory (RMT)\cite{baranger94,jalabert94} the transmission
eigenvalue distribution $P(\lambda)$
has the functional form $P(\lambda)=1/(\pi\sqrt{\lambda(1-\lambda)})$\,,
with a preponderance of eigenvalues near $\lambda=0$ and $\lambda=1$.
Specific features of quantum transport are engraved in the intermediate
values of the $\lambda$-distribution. As it turns out, similar conclusions
apply to the properties of Fano resonances. The values for 
$\overline{T}^{\rm \,offs}_i$\,, plotted as a function of the eigenchannel
number $i$ in ascending order (Fig.~\ref{fig:9}) directly mirrors the
$\cup$-shape of $P(\lambda)$ with a clustering of $\overline{T}^{\rm \,offs}_i$
in the interval $\overline{T}^{\rm \,offs}_i<0.1$ and $\overline{T}^{\rm
    \,offs}_i>0.9$. 
Note that only a few eigenchannel numbers feature
intermediate values of $\overline{T}^{\rm \,offs}_i$. Precisely those
intermediate channels provide the dominant contribution to the resonance
amplitudes $\overline{A}_i$ (Fig.~\ref{fig:9}). This observation suggests a
simple semiclassical explanation: eigenvalues $\lambda$ close to $\lambda=0$
are associated with classical channels of ``pure'' reflection while those
close to $\lambda=1$ are associated with ``pure'' transmission. Classical
transmission or reflection channels correspond to classical path 
bundels\cite{wirt97} of sufficient size in phase space 
($>\hbar$, see Fig.~\ref{fig:5}),
such that a quantum wavepacket can be accommodated.\cite{silve03,jacquod} 
Conversely,
intermediate transmission values correspond to a highly structured area of 
phase space
where the wavepacket encompasses both transmitting and reflecting classical
paths giving rise to quantum indeterminism and 
interferences reflected in $A_i$. Turning now to $q_i$\,, we
find again that the Fano parameter is uncorrelated with
$\lambda_i$\,. Instead, the values for $q_i$ appear to be
``randomly'' distributed (not shown). To quantify their randomness, we plot
the probability
distribution of the magnitude $|q|$, denoted by $P(|q|)$ (Fig.~\ref{fig:10}).
Within the limited statistics available, $|q|$ appears to be a Gaussian
distributed random variable. Since $q$ is a measure for the ratio of the
coupling to the resonant ($r$) scattering channel, 
relative to the non-resonant continuum ($c$), 
\begin{equation}
|q|\propto\frac{|\langle r|t|i \rangle|}{|\langle c|t|i\rangle|}\,,
\label{e:vierpunktdrei}
\end{equation}  
the distribution $P(|q|)$ peaks near $|\langle r|t|i \rangle|\approx 0$,
corresponding to the limit of a ``window'' resonance. 
It should be noted that for individual resonances $|q|$ is not invariant
under the transformation from the mode representation ($T_{mn}$) to 
the eigenchannel representation ($\lambda_i$). This is because $|q|$
depends explicitly on the ratio of the amplitudes 
[Eq.~(\ref{e:vierpunktdrei})] and thus on the matrix elements themselves.
Assuming that
$|\langle c|t|i\rangle|$ is a smooth, weakly varying function across the
resonance, the coupling (or overlap) of the wavefunction of the resonance with
the entrance $(|i\rangle)$ channel function can be considered to be a Gaussian
random number. Such a hypothesis would agree with RMT
predictions for chaotic wavefunctions\cite{mehta} even though there is no
a-priori reason for the applicability of RMT to the present hierarchical phase
space structure.
\section{Can fractal conductance fluctuations be observed?}\label{sec:fractal}
 Our calculation extends to mode numbers up to $M=40$ thereby reaching a ratio
 of $\lambda_F/D \lesssim 0.01$. Our calculation reaches therefore further
 into the semiclassical regime than previous calculations for 2D billiards. It
 is therefore tempting to probe for the occurrence of fractal conductance
 fluctuations (FCF) in the transmission probability. According to a
 semiclassical argument\cite{ketzprb96} quantum dots with mixed classical
 dynamics can be expected to give rise to self-similar fluctuations (over
 several orders of magnitude) to which a fractal (i.e.~non-integer) dimension
 can be attributed. Previous numerical results failed to provide unambiguous
 evidence for the presence of FCF in these
 systems.\cite{hucke00,hucke00e,baeck02,taka00} This is due to the limited
 values of $k_F$ that could be reached computationally. This difficulty can be
 circumvented by reducing the two-dimensional scattering devices to effectively
 one-dimensional systems, 
 so called quantum graph models. Here numerical constraints
 are less severe thus allowing to explore a regime where both isolated
 resonances and FCF simultaneously exist.\cite{hufnagelepl01} Our present
 two-dimensional calculations for transport through the soft-wall stadium
 (Fig.~\ref{fig:1}) do not show FCF, even for the highest $k_F$ that was
 accessible by our codes, i.e., $k_F \approx 40 \times \pi/d$ or
 $\lambda_F/D\approx 0.01$. 
 We attribute the absence of FCF to the fact that even such $k_F$ are
 still too small to probe the turnstiles of the cantori in classical phase
 space. This statement is supported by our observation that the HDs enter the
 outermost partial barrier only at resonance energies, i.e.~by
 tunneling. Transport into the inner regions of the hierarchy in phase space
 seems to be suppressed by the partial barrier. This exploration of
 hierarchical phase space by way of turnstiles in cantori is, however, crucial
 for the fractal fluctuations.\cite{ketzprb96}\\Our present negative result
 for soft-wall billiards even for moderately large $k_F$ leaves, however, the
 question open which mechanism is at work that has apparently produced
 signatures of FCF in several recent
 experiments.\cite{sachr98,micolich01,micolich02,micolich04,crook03} 
These experiments were in
 the regime where only a rather limited number of transmitting modes $m, n = 2
 - 6$ is open.\cite{micolich01} Our results clearly show that the 
presence of soft walls in the experiment
 can be ruled out as the source for FCF at moderate $k_F$. Note that this
 observation is in close correspondence to recent findings which
 suggest that ``more complicated processes than those predicted
 in the
 semiclassical models are responsible for the observed behavior of 
 FCF''.\cite{micolich04}
 \section{summary}\label{sec:summary}
 We have investigated the classical and quantum scattering properties for a
 soft-wall billiard with mixed phase space representing the generic device
 used in experimental realizations. By analyzing the wave function probability
 density and the Husimi distribution of scattering states we find remarkable
 similarities between the classical and quantum phase space structures. This
 enables us to classify resonant scattering states associated with regular,
 trapped and instable periodic classical trajectories. Such a mapping of
 resonant scattering states is mirrored in characteristic differences in 
 the width
 of the corresponding resonances. Our investigations reveal that the observed
 resonances in all the partial transmission amplitudes $T_{mn} (k_F)$ follow
 the asymmetric Fano lineshape. The distribution of Fano asymmetry parameters
 $q$ appears to be surprisingly uncorrelated with the injection and ejection
 angles of the classical trajectories. However, the resonance amplitude is
 approximately proportional to the geometric mean of the lead mode numbers
 $\sqrt{m \times n}$. Studying the transmission eigenvalues $\lambda_i$ 
 of $\hat{T}=t t^\dagger$, we find that Fano resonances in $\lambda_i$
 feature $q$-parameters following a Gaussian distribution and amplitudes 
 $A_i$ that have substantial contributions only in the non-classical 
 transmission eigenchannels.\cite{silve03}
 For numerically accessible wavenumbers with $\lambda_F/D
 \approx 0.01$ fractal conduction fluctuations (FCF) could not be detected.
\begin{center}
{\bf Acknowledgments }
\end{center}
We thank A.~B\"acker, L.~Hufnagel, F.~Libisch, and R. Ketzmerick
for helpful discussions. Support by the Austrian Science
Foundation (Grant No.~FWF-P17359 and No.~FWF-P15025) is gratefully
acknowledged. 
\bibliographystyle{apsrev}
%\bibliography{mybibliography}

\end{document}